\documentclass[12pt]{article}
\usepackage{fullpage}
\usepackage{amssymb, amsthm, amsmath}
\usepackage{doublespace}
\usepackage{graphicx}
\usepackage[authoryear]{natbib}
\usepackage{bm}
\usepackage{verbatim}
\usepackage{lineno}
\usepackage{times}
\usepackage{tabulary}
\usepackage{url}
\usepackage{gensymb}

\newcommand{\bbeta}{ \mbox{\boldmath $\beta$}}

\newcommand{\blambda}{ \mbox{\boldmath $\lambda$}}

\newcommand{\bX}{ \mbox{\bf X}}

\newcommand{\bs}{ \mbox{\bf s}}

\newcommand{\beq}{ \begin{equation}}
\newcommand{\eeq}{ \end{equation}}
\newcommand{\beqn}{ \begin{eqnarray}}
\newcommand{\eeqn}{ \end{eqnarray}}

\newcommand{\est}{\widehat}
\newcommand{\PM}{ \mbox{PM$_{2.5}$}}

\begin{document}

\begin{center}
{\Large A comparison of statistical and machine learning methods for creating national daily maps of ambient $\PM$ concentration}\\\vspace{6pt}
{\large Veronica J. Berrocal\footnote{Department of Biostatistics, School of Public Health, University of Michigan}, Yawen Guan\footnote{Department of Statistics, North Carolina State University}\footnote{The Statistical and Applied Mathematical Sciences Institute}, Amanda Muyskens$^2$,
Haoyu Wang$^2$, Brian J Reich$^2$,  James A. Mulholland\footnote{Environmental Engineering, Georgia Tech} and
Howard H. Chang\footnote{Department of Biostatistics and Bioinformatics, Rollins School of Public Health, Emory University}
}\\
\today
\end{center}
\begin{abstract}\begin{singlespace}
\noindent 
A typical problem in air pollution epidemiology is exposure assessment for individuals for which health data are available. Due to the sparsity of monitoring sites and the limited temporal frequency with which measurements of air pollutants concentrations are collected (for most pollutants, once every 3 or 6 days), epidemiologists have been moving away from characterizing ambient air pollution exposure solely using measurements. In the last few years, substantial research efforts have been placed in developing statistical methods or machine learning techniques to generate estimates of air pollution at finer spatial and temporal scales (daily, usually) with complete coverage. Some of these methods include: geostatistical techniques, such as kriging; spatial statistical models that use the information contained in air quality model outputs (statistical downscaling models); linear regression modeling approaches that leverage the information in GIS covariates (land use regression); or machine learning methods that mine the information contained in relevant variables (neural network and deep learning approaches). Although some of these exposure modeling approaches have been used in several air pollution epidemiological studies, it is not clear how much the predicted exposures generated by these methods differ, and which method generates more reliable estimates. In this paper, we aim to address this gap by evaluating a variety of exposure modeling approaches, comparing their predictive performance and computational difficulty. Using PM$_{2.5}$ in year 2011 over the continental U.S. as case study, we examine the methods' performances across seasons, rural vs urban settings, and levels of PM$_{2.5}$ concentrations (low, medium, high).
\vspace{12pt}\\
{\bf Key words:} Air pollution exposure assessment; spatial modeling; kriging; Bayesian hierarchical modeling; downscaling; machine learning; random forest; support-vector regression; neural network. 
\end{singlespace}\end{abstract}
\newpage

\section{Introduction}\label{s:intro}
Accurate exposure assessment plays an essential role in the success of any environmental health study. Past air pollution epidemiological studies regularly utilize ambient air quality measurements from large monitoring networks to estimate population and individual exposures. However, measurements from these networks are spatially sparse, temporally incomplete, and preferentially located in settings with large populations and high pollution levels. There is increasing interest in developing methods to retrospectively estimate air pollution levels at fine spatial scales and with complete spatial-temporal coverage in order to minimize exposure measurement error \citep{alexeeff2015consequences}, support more spatially-resolved health effect analyses \citep{kloog2012acute, hao2016air}, and perform impact assessments in low- and middle-income settings \citep{shaddick2018}. Advances in Geographical Information Systems, remote sensing, and numerical model simulations have further contributed to a proliferation of modeling approaches to estimate air pollution over the past decade. 

This paper aims to address an important gap in the current literature. When models to estimate ambient air pollution exposure are being developed, they are typically only compared to simpler models within the same modeling paradigms. Examples include  comparisons between regression models with different predictors, especially in land use regression models \citep{tang2013using, wang2016combining}; between geostatistical models with different spatial dependence structures \citep{reich2011class}; or between different algorithms among the machine learning toolbox \citep{singh2013identifying, reid2015spatiotemporal}. There has been very limited cross-paradigm comparisons \citep{adam2014spatiotemporal, yu2018cross}, likely due to the analytic effort and expertise required to carry out the different approaches. In particular, there is a lack of comparison between machine learning and advanced geostatistical approaches. As exposure modeling becomes increasingly complex and computationally demanding, there is a pressing need to better understand the advantages and limitations associated with different modeling approaches. However, synthesis of results from existing studies is challenging due to the use of different data sources and inconsistent criteria for evaluating model performance. 

The main objective of this paper is to critically evaluate different exposure modeling approaches via a case study of estimating regional daily ambient fine particulate matter of aerodynamic diameter less than 2.5 $\mu m$ ($\PM$ ) across the contiguous United States. We considered major classes of recent approaches that can provide complete spatial and temporal coverage using monitoring measurements, meteorology, land use variables, and numerical model simulations. These approaches include geographically weighted regression\citep{van2015high} , statistical downscaling \citep{berrocal2010spatio}, and machine learning methods, including random forest \citep{hu2017estimating}, support vector machine \citep{liu2017urban}, and neural networks \citep{di2016assessing}. Air quality estimates from these approaches have already been used in health effect and health impact analyses \citep{chang2011time, strickland2016pediatric, di2017association}. Our evaluations focuses specifically on a model's spatial predictive ability at locations without monitors and on computational demand.    

A related issue that has hindered cross-comparison is the limited availability of data and analytic code from exposure modeling studies. This is in contrast to efforts in other fields, such as genetics, genomics and neuroimaging, where established publicly available and well-documented datasets are available for methods development and comparison \citep{mailman2007ncbi, leinonen2010sequence, van2013wu}. Sophisticated exposure models and algorithms often require considerable efforts in input associated with data processing and parameter tuning to maximize model performance. A notable example is the air pollution modeling framework for 2-week averages from the MESA Air study \citep{keller2015unified} where an R package has been made available \citep{lindstrom2012spatiotemporal}. Hence, another contribution of this study is the documentation of all data and analytic code. We also only considered methods that can be readily implemented using packages from the statistical R software \citep{R}.  This will ensure better reproducibility, allow for future evaluation of other methods, and facilitate adaptation of these modeling methods by other research groups.

\section{Data sources}\label{s:data}

\subsection{$\PM$ Monitoring data}\label{s:data:monitors}

\begin{figure}[hpb]\centering
\includegraphics[width=0.48\textwidth]{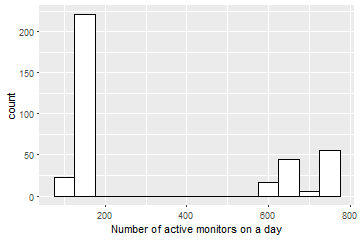}
\caption{{\bf Number of active monitors on a day.}}\label{f:num_monitors}
\end{figure}

For year 2011, daily $\PM$ measurements were obtained from the Air Quality System database of the US Environmental Protection Agency. A total of 829 monitors from state, local, and tribal air pollution monitors are available across our study domain of the contiguous United States.  Monitors typically take measurement every 3 or 6 days, with only 15\% of monitors sampling almost daily. Figure \ref{f:num_monitors} shows a histogram of the number of active monitors each day: most days have less than 200 measurements across the contiguous US.  \\

\begin{figure}\centering

\includegraphics[page=1,width=0.48\textwidth,trim={0 3cm 0 3cm},clip]{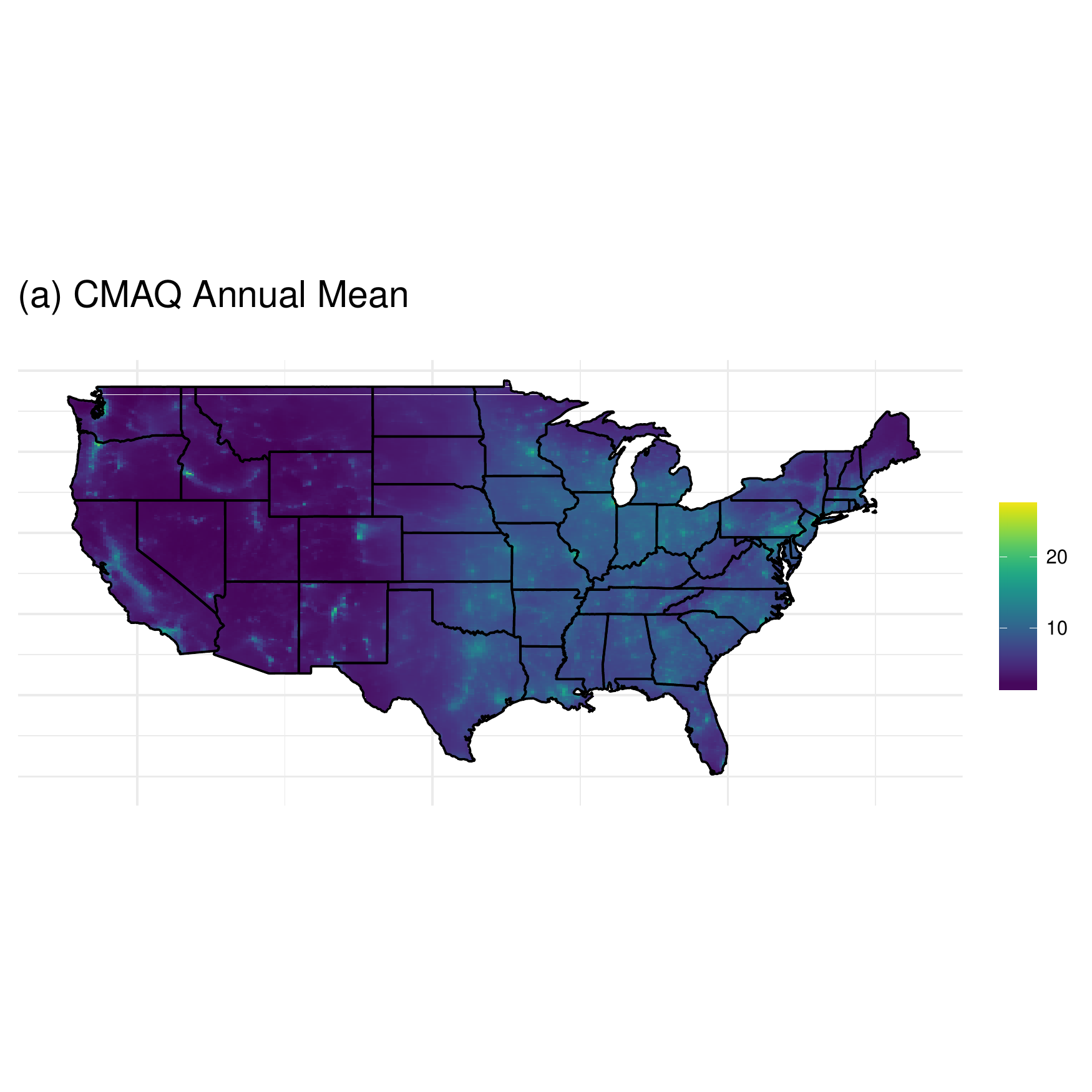}
\includegraphics[page=2,width=0.48\textwidth,trim={0 3cm 0 3cm},clip]{figs/EDA.pdf}
\caption{{\bf Annual mean $\PM$ concentration for year 2011} as estimated by CMAQ output versus AQS monitor data.}\label{f:monitors}
\end{figure}

\subsection{Computer model output}\label{s:data:cmaq}

Computer models for air quality use information on emission sources to simulate the fate and transport of air pollutants by taking in account complex atmospheric chemistry and physics. There is increasing use of computer model outputs to estimate air pollution levels at locations without monitoring data \citep{berrocal2010spatio}. We obtained Community Multiscale Air Quality model \citep[CMAQ;][]{CMAQ} version 4.7 runs performed by the EPA at 12km spatial resolution across the contiguous US for year 2011. At this resolution, the contiguous US is covered by a grid of 299 by 459 cells. Each $\PM$ monitor was linked to the CMAQ grid cell that contains it.  Figure \ref{f:monitors} compares annual average $\PM$ concentrations obtained from CMAQ simulations and AQS measurements.

\subsection{Meteorological and land use data}\label{s:data:met}
Daily average meteorological fields were obtained from the North American Land Data Assimilation System (NLDAS) and the North American Regional Reanalysis (NARR). Meteorological variables include temperature at 2m, 30m, 150-180mb above ground, downward longwave and shortwave radiation flux, relative humidity at 2m, boundary layer height, visibility, dew-point temperature, potential evaporation,  potential evaporation, convective available potential energy, pressure at 2m, 10m, and 30m, and U- and V-wind speed at 10m. We also obtained the following land use variables: elevation from the US Geological Survey, major roadway lengths and percentage of forest cover from the National Land Cover database, population density from the U.S. 2010 Census Bureau at the census tract level, percentage of impervious surface from Landsat,  and $\PM$ primary emission sources from the 2011 US EPA National Emissions Inventory Facility Emissions report.  Although all the above mentioned variables were available at different spatial resolution (point level, in some cases, grids, etc.), they were re-gridded to the CMAQ 12km grid. Specifically, for predictors with finer spatial resolutions than CMAQ (e.g. elevation), we used the average of cell values that intersect with the CMAQ cell. For predictors with coarser spatial resolution than CMAQ (e.g. meteorology), the value of the nearest grid cell centroid was used. 

\subsection{Variable selection}\label{s:var_selection}
It is well established that meteorological variables and land use characteristic are good predictors of $\PM$ concentration, as land use regression models indicate \citep{hoek2008}. To identify which of the available predictors should be used in the exposure models considered in our subsequent model comparisons, we first performed variable selection via cross validation and best subset regression \citep{kutner2005applied}. For this, we randomly split the monitors into 5 folds. For each fold, we performed best subset regression using data from the other 4 folds and  predicted $\PM$ concentration at the hold-out fold. For each hold-out set, we computed the root mean square error (RMSE), comparing the predicted values with the held-out $\PM$ observations. We averaged the RMSEs across the five folds and selected the model which yielded the best predictive performance. This identified a model with 11 predictors:  including an additional predictor only improved prediction by less than 0.1\%. After having determined the number of predictors to include in the model, we identified the set of 11 predictors again via best subset regression. In this second step, best subset regression was performed on the full data using all meteorological variables and land use covariates. The 11 selected predictors are shown in Table \ref{tab:varselection}. The set consisted mostly of meteorological variables, likely due to our interest in estimating daily $\PM$ concentrations.  

\begin{table}
\begin{center}
\caption{Set of eleven predictors selected via best subset regression. In parentheses it is indicated the original database from which each predictor was obtained: North American Land Data Assimilation System (NLDAS) and North American Regional Reanalysis (NARR).}\label{tab:varselection}
\begin{tabular}{|l|l|}
\hline\hline
Percentage of Impervious Surface (Landsat) & Potential Evaporation (NLDAS) \\
Downward Longwave Radiation Flux (NLDAS) & Convective Available Potential Energy (NLDAS)\\
Relative Humidity (NLDAS) & V-Wind at 10m (NLDAS) \\
Pressure (NLDAS) & Dew-point Temperature (NARR) \\
Visibility (NARR) & Boundary Layer Height (NARR) \\
Temperature 150-180mb above ground (NARR) \\
\hline
\end{tabular}
\end{center}
\end{table}

\section{Methods}\label{s:methods}
\subsection{Statistical methods}\label{s:stats}

Let $Y_t(\bs)$ be the monitor measurement of $\PM$ at spatial location $\bs$ and day $t$.  Each observation is associated with land-use covariates and meteorological variables at $\bs$, $\bX_t(\bs)$, and the CMAQ output at the grid cell that contains $\bs$, $Z_t(\bs)$.  We assume monitoring data are available at $n$ sites $\bs_1,...,\bs_n$ and for days $t=1,...,T$. The distance (in km) between location $\bs_i$ and $\bs_j$ is denoted $d_{ij}$.  The objective of all methods is to make a prediction $\est{Y}_t(\bs_0)$ of $\PM$ concentration at a location $\bs_0$ that does not have a monitor, with prediction uncertainty quantified through $\mbox{Var}[\est{Y}_t(\bs_0)] = \est{v}_t(\bs_0)$.      

\subsubsection{Inverse distance weighting}\label{s:stats:grw}

Inverse distance weighting (IDW) does not use the covariates and simply uses a weighted average of nearby observations as the prediction.  The intuition behind IDW is that the weights for observations near the prediction location should be higher than the weights for observations far from the prediction location.  The weight assigned to the observation at location $\bs_i$ for prediction at location $\bs_0$ decays as the distance $d_{0i}$ between $\bs_0$ and $\bs_i$ increases,

\begin{equation}\label{e:GWR:W}
  w_{0i} = \left(\frac{1}{d_{0i}}\right)^{\phi},
\end{equation}
where the rate of decay is controlled by $\phi>0$, which we select using cross validation. 

The IDW prediction is 
\begin{equation}\label{e:GWR:IDW}
  \est{Y}_t(\bs_0) = \sum_{i=1}^nW_{0i}Y_t(\bs_i)
  \mbox{\ \ \ where \ \ \ }
  W_{0i} = \frac{w_{0i}}{\sum_{j=1}^nw_{0j}}.
\end{equation}
The weights $W_{0i}$ are standardized to sum to one so that predictions are unbiased.  For large $n$, computational speed can be increased by setting $w_{0i}=0$ for locations with large $d_{i0}$'s to reduce the number of terms in the sum in (\ref{e:GWR:IDW}).  Because no statistical model is assumed, there is no natural variance estimate, $v_t(\bs)$.

\subsection{Geostatistical methods}\label{s:stats:krig}

In geostatistics, the problem of generating predictions of a continuous spatial process at unsampled locations is typically handled via Kriging. As with IDW, in Kriging the prediction $\hat{Y}_{t}(\bs_0)$ is a linear combination of the observations $Y_{t}(\bs_1), \ldots, Y_{t}(\bs_n)$ with weights $\blambda=(\lambda_1,\lambda_2,\ldots, \lambda_n)$, that is, $Y_{t}(\bs_0) = \sum_{i=1}^n \lambda_i Y_t(\bs_i)$.  The weight vector $\blambda$ is in turn determined by finding the set of $\lambda$'s that yield the Best Linear Unbiased Predictor (BLUP), e.g.
$$
\blambda = \mbox{arg min}_{{\tilde \blambda}} E \left[ Y_{t}(\bs_0) - \sum_{i=1}^n {\tilde \lambda}_i Y_t(\bs_i) \right]^2 \qquad \mbox{such that} \qquad E\left[\sum_{i=1}^n {\tilde \lambda}_i Y_t(\bs_i)\right]=E \left[Y_{t}(\bs_0) \right]
$$
The expression for the weights depends on the assumptions made on $Y_t(\bs)$, such as whether the process is second-order stationary (i.e., the covariance depends only on the separation between points) and whether its mean varies spatially and depends on covariates. For computational tractability, we assume a stationary covariance function \citep{cressie1993statistics}. In addition, given that we are generating predictions of $\PM$ concentration across the entire United States it is safe to assume that the mean concentration varies spatially as function of covariates. Thus, we will generate our predictions using Universal Kriging \citep{cressie1993statistics}. The latter, differently from Ordinary Kriging \citep{cressie1993statistics}, includes covariates in the expression of the mean function.

\subsubsection{Universal Kriging}
In Universal Kriging we assume that, for each day $t$, $Y_t(\bs)$ is a Gaussian process with a pre-determined covariance function and a mean $\text{E}[Y_t(\bs)]=\mu_t(\bs)$, linear function of spatially-varying covariates. In our implementation, as we are dealing with data over time, we assume that the spatial covariance does not change in time and the correlation between $\PM$ concentration at two sites decays exponentially with distance.  In other words, we use a stationary exponential covariance function with covariance parameters that are constant in time \citep{cressie1993statistics}. We compare different scenarios for $\mu_t(\bs)$:
\begin{enumerate}
\item $\mu_t(\bs)$ depends only on the CMAQ output $Z_t(\bs)$, e.g.
\begin{equation}
\mu_t(s) = \gamma_{0,t} + \gamma_{1,t} Z_t(\bs)
\label{uk:onlyCMAQ}
\end{equation}
\item $\mu_t(\bs)$ depends only on meteorological and land-use covariates $\bX_{t}(\bs)$, e.g.
\begin{equation}
\mu_t(s) = \bX_t(\bs) \bbeta_t
\label{uk:onlycovs}
\end{equation}
\item $\mu_t(\bs)$ depends on both the CMAQ output $Z_t(\bs)$ and the meteorological and land-use covariates $\bX_{t}(\bs)$, e.g.
\begin{equation}
\mu_t(s) = \bX_t(\bs) \bbeta_t +  Z_t(\bs)\gamma_{1,t}  
\label{uk:CMAQandcovs}
\end{equation}
\end{enumerate}
To implement the different universal Kriging models, we use a two-stage procedure and use the \texttt{gstat} and \texttt{geoR} \citep{geoR} packages in \texttt{R} \citep{R}. Specifically, we first estimate the spatial covariance parameters via weighted least squares (WLS) using the \texttt{gstat} package, fitting an exponential semi-variogram to the empirical semi-variogram of the time-averaged residuals of the linear regressions implied by (\ref{uk:onlyCMAQ}), (\ref{uk:onlycovs}) and (\ref{uk:CMAQandcovs}), respectively. 
Using the spatial covariance parameters estimated via WLS as initial values, we used the \texttt{geoR} package \citep{geoR} and maximum likelihood to iteratively estimate the regression parameters parameters $\bbeta_t$, $\gamma_{0,t}$ and $\gamma_{1,t}$ and the covariance parameters. Using such estimates, we then generate predictions of $Y_t(\bs_0)$ via Universal Kriging separately for each day. Because the data are assumed to be Gaussian, Kriging predictions are  accompanied by predictions variance estimates ${\hat v}_t(\bs)$.

\subsection{Downscaler}\label{s:stats:SD}
The downscaler model generates predictions of air pollution concentration at any location $\bs$ by exploiting the relationship between the observed concentration measured at a monitor and the estimated air pollution concentration generated by an air quality model, CMAQ. The simplest downscaler model \citep{berrocal2010spatio} relates the (log or square root of the) observed concentration at location $\bs$ on day $t$, $Y_t(\bs)$, to the CMAQ output, $Z_{t}(\bs)$, at the grid cell that contains location $\bs$ via a linear regression model with spatially and temporally-varying coefficients. Choosing the form of the downscaler model that yields the best predictive performance in various experiments, we adopt the following version of the downscaler model 
\begin{equation}
Y_{t}(\bs) = \beta_{0,t}(\bs) + \beta_{1,t} Z_t(\bs) + \epsilon_{t}(\bs) \qquad \epsilon_{t}(\bs) \stackrel{iid}{\sim} N(0,\tau^2)
\label{eq:spatial_reg}
\end{equation}
where $\beta_{0,t}(\bs)$ is a spatially and temporally-varying intercept term, while $\beta_{1,t}$ indicates a slope term, constant in space but varying in time.

As on a given day $t$, we postulate that sites located nearby have a similar intercept term, to account for spatial dependence, we model $\beta_{0,t}(\mathbf{s})$ as a stationary spatial processes with an exponential correlation function. That it, for any two pair of sites $\bs_i$ and $\bs_j$: 
\begin{equation}
\mbox{Cov} \left( \beta_{0,t} (\bs_i), \beta_{0,t}(\bs_j) \right) = \sigma^2_0 \exp \left( - \frac{d_{ij}}{\phi_0} \right)
\label{eq:cov_fct}
\end{equation}
In (\ref{eq:cov_fct}), $d_{ij}$ denotes the distance between sites $\bs_i$ and $\bs_j$, while $\sigma^2_0$ and $\phi_0$ represent, respectively, the spatial variability in $\beta_{0,t}(\mathbf{s})$, and the rate at which the spatial correlation vanishes.

The downscaler model in (\ref{eq:spatial_reg}) is fit within a Bayesian framework \citep{gelman2013bayesian}, and thus its specification is completed once prior distributions for all the model parameters are provided. Specifically, we assume that, for each day $t$, the spatially varying intercept $\beta_{0,t}(\mathbf{s})$ admits a constant mean $\beta_{0,t}$, which is interpreted as the overall additive calibration of the CMAQ output. In contrast, $\beta_{1,t}$ represents the overall multiplicative calibration term for CMAQ. The two global calibration parameters $\left( \beta_{0,t} \; \beta_{1,t} \right)^\prime$ are assumed to be independent in time and follow a bivariate normal distribution with mean $\left( 0 \; 1 \right)^\prime$ and with a diagonal, covariance matrix with large prior variances. The two variance parameters, $\tau^2$ and $\sigma^2_0$, representing, respectively, the non-spatial and spatial variability in $\PM$ concentration, are provided with vague Inverse Gamma priors, while Uniform prior on the interval $(0.0001 km, 0.1 km)$ is placed on the decay parameters $\phi_0$.

Inference on model parameters is carried out through the posterior distribution, which is approximated using an MCMC algorithm \citep{gelman2013bayesian} we ran for 10,000 iterations, with the first 5,000 discarded for burn-in. For the case study discussed in the paper, we  make a slight modification to the downscaler model, allowing the spatial dependence parameters ($\sigma^2$, $\phi$) and the non-spatial variance $\tau^2$ to vary with time. With this modification we can implement the downscaler model, using the \texttt{spBayes} \citep{spbayes} package in \texttt{R}.

Predictions of air pollution concentrations on day $t$ at an unsampled location $\bs_0$ and uncertainty estimates for the predictions are obtained using the posterior predictive distribution of $Y_{t}(\bs_0)$ given the observed data. Specifically: we take as predicted concentration, the median of the posterior predictive distribution, while we characterize uncertainty in the prediction via the $95\%$ predictive interval, whose upper and lower limits are given by the 2.5-th and 97.5-th percentile of the posterior predictive distribution. Finally, we use the sample variance ${\hat v}_t(\bs)$ of the predicted values to quantify the variance in the predictions.

\subsection{Machine learning methods}\label{s:stats:ML}

The spatial regression models in Section \ref{s:stats:krig} and \ref{s:stats:SD} represent the data-generating process with a small number of interpretable parameters.  In contrast, machine learning algorithms have countless uninterpretable parameters and are thus essentially black-box prediction machines.  However, these algorithms are flexible, general and have often outstanding predictive performance.  Below we briefly describe the machine learning algorithms used in our comparison; for a detailed description of these and other machine learning algorithms see \cite{james2013introduction}.

For application of machine learning methods we do not explicitly model spatial correlation, although spatial coordinates and spatial covariates are used as predictors in these regression models.  Since we do not model spatial correlation we denote the observed $\PM$ concentration and covariates relative to observation $i=1,...,n$ ($n$ is the combined sample size over space and time) as $Y_i$ and $\bX_i$, respectively.  The $p=15$ covariates in $\bX_i$ are the longitude, latitude, day of year, the covariates in Table \ref{tab:varselection} and CMAQ model output.

\subsubsection{Random forests}

A random forest \citep{Breiman2001} is an ensemble of regression trees \citep{Morgan_Sonquist1963}.  A regression tree is grown via recursive partitioning of the covariate space called ``leaves", and then fitting separate linear models within each leaf. For each split of the covariate space, $m$ input variables are selected at random as candidates for splitting. Randomizing over the covariates decreases the correlation between trees and  improves the prediction accuracy of the ensemble. The intuition is that interactions are likely present when considering the entire covariate space, but when considering small subregions of the covariate space simple linear models likely fit well within each subregion.  

To construct a random forest for $\PM$ concentration, many regression trees are grown with the prediction being a weighted average of the predictions over the trees.  
Unlike most machine learning methods, it is possible to quantify the uncertainty in the prediction $v_t(\bs)$ using the sample variance of the trees predictions. Random forest is a desirable data mining method because it is easily understood and it is computationally efficient for very large samples. The number of trees used for prediction is selected by out-of-bag error \citep{hastie01statisticallearning}, a cross validation technique where the observations that are not selected in a tree are then used to estimate the error. For our case study, we implemented random forest using the \texttt{R} package \texttt{randomForest} \citep{RF}. We used the default values $m = p/3$ at each split, where $p=15$ is the number of covariates. Based on out-of-bag error, a random forest with 500 trees is used as the final predictive model.

\subsubsection{Support Vector Regression (SVR)}

Nonlinear support vector machines \citep{cortes1995support} are most well known in machine learning as a classification tool, but support vectors can also be used for regression. In SVR a hyperplane is optimized to be within a certain threshold of the selected data, called the support vectors, and the hyperplane is used for regression prediction.  The predicted $\PM$ concentration for an observation with covariates $\bX_0$ is $$\hat{Y}_0 =  \sum_{i=1}^{n}K(\bX_i,\bX_0)(\alpha_i - \alpha_i^{\star})$$
where $\alpha_i$ and $\alpha_i^{\star}$ are the support vectors and $K(\bX_i,\bX_j)=\exp(-|\bX_i-\bX_j|^2/\gamma)$ is the radial basis kernel function.  The support vectors $\alpha_i$ and $\alpha_i^*$ are estimated as the solution to the convex optimization
\begin{equation}
\frac{1}{2} \sum_{i=1}^n \sum_{j=1}^n (\alpha_i - \alpha_i^{\star})^{T} (\alpha_j - \alpha_j^{\star}) K(\bX_i, \bX_j) + \epsilon \sum_{i=1}^{n}(\alpha_i + \alpha_i^{\star}) + \sum_{i=1}^{n}Y_i(\alpha_i - \alpha_i^{\star})
\label{eq:svm}
\end{equation}
so that for all $i$, $0<\alpha_i<C$,   $0<\alpha_i^{\star}<C$ and $ \sum_{i=1}^{n}(\alpha_i - \alpha_i^{\star})=0$. The constant $C$ is the box constraint that defines the trade off between penalty for observations further than $\epsilon$ away from the hyperplane and smoothness in the prediction hyperplane.


This type of model is known to yield good predictions, especially in high dimensional space. However, we note that the calculation of the optimum solution to (\ref{eq:svm}) is computationally expensive in large datasets and there is no straightforward method to estimate prediction variances. We implement support vector regression using the R package {\tt e1071} \citep{dimitriadou2006e1071} and the function {\tt svm}.

\subsubsection{Neural networks}
Neural networks are machine learning methods that have successfully been used to predict air pollution levels \citep{di2016assessing}. They are attractive because they can handle massive data and can model nonlinearity and interactions. We use a multilayer perceptron network \citep{rumelhart:errorpropnonote} that consists of an input layer, several hidden layers and an output layer.  For an observation with covariates $\bX_0=(X_{01},...,X_{0p})$ the predicted value with a single hidden layer is 
\begin{eqnarray}
  \mbox{Output layer}:  && {\hat Y}_0= b_1^3 + \sum_{j=1}^{L_2}W_{j1}^3Z_{0j}^2\\
  \mbox{Hidden layer}: && Z_{0l}^2 = f\left(b_l^2+\sum_{j=1}^{L_1}W_{jl}^2Z_{0j}^1\right) \mbox{ for } l\in\{1,...,L_2\}\nonumber\\
  \mbox{Input layer}: && Z_{0l}^1 =  f\left(b_l^1+\sum_{j=1}^{p}W_{jl}^1X_{0j}\right) \mbox{ for } l\in\{1,...,L_1\},\nonumber
\end{eqnarray}
where $L_1$ and $L_2$ are the number of neurons in each layer, $f$ is the activation function, and $b_l^k$ and $w_{jl}^k$ are the bias and weights estimated to minimize mean squared error in the training data. 

Fitting a neural network requires tuning the number of hidden layers, the numbers of neurons in each layer, and the activation function. Increasing the size of the network provides more flexibility but also may lead to overfitting. Therefore, we also tune the number of epochs, drop-out rate, learning rate, and minibatch size \citep{Goodfellow2016}.  The model is fit using the deep learning package {\tt keras} \citep{keras} in \texttt{R}. We tried relu and sigmoid activation functions and networks with 1 to 3 hidden layers, with 500 to 2000 neurons in the the first layer, and a reduced number of neurons in each of the next layers. Further tuning with the following possible combinations were considered: number of epochs ranging from 50 to 200, drop-out rate ranging from 0.1 to 0.5, learning rate ranging from 0.0005 to 0.01, and minibatch size ranging from 128 to 1024. Based on five-fold cross-validation, the final selected model has 2 hidden layers with 2000 and 100 neurons for the first and second layer, respectively, relu activation function $f(x)=\text{max}\{0,x\}$, 200 epochs, 0.4 drop-out rate, 0.003 learning rate and 1024 minibatch size.

\section{Results}\label{s:results}

We compare the statistical and machine learning methods described in Section~\ref{s:methods} by assessing their predictive performance. We use five-fold cross validation, and randomly sample, without replacement, 20\% of the sites to be in the test set in each of the five folds. Within each fold, we train each of the models on the data from the remaining 80\% of the sites and we generate predictions at the test sites.  We evaluate the predictive performance of the various methods in terms of root mean squared error (RMSE), mean absolute deviation (MAD), correlation between predicted and observed $\PM$ concentrations, and coverage of the 95\% prediction intervals, all averaged across days and sites. Table~\ref{t:daily:PM} presents the results.

\begin{table}\caption{{\bf Cross-validation results of daily $\PM$ predictions:} We compare the various statistical and machine learning methods using root mean squared error (``RMSE''), mean absolute deviation (``MAD''), correlation between predicted and observed values (``Corr''), and empirical coverage of the predictive 95\% intervals.  Both RMSE and MAD are in units of $\mu g/m^3$. The methods applied are ordinary least squares (``OLS''), inverse distance weighting (``IDW''), universal Kriging (``UK''), downscaler, random forests (``RF''), support vector regression (``SVR'') and Neural networks (``NN'').  Methods use either CMAQ and/or other geographic covariates (``Covs'').}\label{t:daily:PM}	
\begin{center}\begin{tabular}{l|cccc}
Method & RMSE & MAD & Corr & Coverage\\\hline 
OLS (CMAQ)         & 4.80 & 3.09 & 0.65 & 0.62 \\
OLS (Covs)         & 4.63 & 2.97 & 0.68 & 0.79 \\
OLS (CMAQ + Covs)  & 4.22 & 2.63 & 0.74 & 0.83 \\
IDW                & 3.39 & 1.96 & 0.84 & -- \\
UK (CMAQ)          & 3.08 & 1.90 & 0.87 & 0.95 \\
UK (Covs)          & 3.25 & 1.79 & 0.85 & 0.93 \\
UK (CMAQ + Covs)   & 3.15 & 1.76 & 0.86 & 0.93 \\
Downscaler (CMAQ)  & 3.10 & 1.70 & 0.87 & 0.94 \\
RF (CMAQ + Covs)  & 3.41 & 2.09 & 0.84 & 0.96\\
SVR (CMAQ + Covs)  	& 3.83 & 2.22 & 0.79 & --\\
NN (CMAQ + Covs)   	& 3.89 & 2.45 & 0.79 & --\\
\end{tabular}\end{center}\end{table}

Predictions obtained via a linear regression model fit separately to $\PM$ concentration each day serves as a baseline for comparison, and it yields an RMSE of 4.22$\mu g/m^3$ when both CMAQ and the selected covariates listed in Table~\ref{tab:varselection} are included in the model.  Machine learning methods yield predictions that are substantially improved over those obtained via least squares, by accounting for potential non-linear and interaction effects of the covariates.  Of the machine learning algorithms, Random Forest has the smallest RMSE (3.41$\mu g/m^3$).  All of the statistical methods that explicitly model the spatial correlation in the data have smaller RMSE than the machine learning methods.  Inverse distance weighting which generates prediction by simply taking a local average of the available monitoring data has an RMSE of 3.39$\mu g/m^3$. The best method in terms of RMSE is Universal Kriging (UK) when only CMAQ is used as a covariate.  The spatial downscaler model yields a comparable RMSE (3.10$\mu g/m^3$ vs 3.08$\mu g/m^3$) but has a smaller MAD.

Table~\ref{t:nclose} examines whether the quality of the predictions is improved based on the number of nearby monitoring sites with data available on the day of the prediction. The table presents predictive performance results stratified by the number of monitoring stations within a 50 miles radius of the prediction location with same-day PM$_{2.5}$ measurements.  As with overall predictions, the geostatistical methods have the smallest prediction error for all numbers of active training sites.  For test sites with many active neighbors, the downscaler model has slightly smaller prediction error than UK, but UK performs well in all cases.  The Appendix includes analogous tables that compare the quality of the predictions generated by the various methods when we stratify by distance to nearest monitoring location, longitude, and season, while this section presents results stratified by urbanization (Table \ref{tab:by_urbanization}) and  PM$_{2.5}$ magnitude (Table \ref{tab:by_magnitude}).  For all methods, prediction errors are smaller at urban sites, sites near other monitoring stations, sites in the Eastern US and sites with low PM$_{2.5}$ concentrations.  Prediction errors are also lower in the Spring and Summer than in the Fall and Winter.  However, the ranking of the prediction methods remains similar across all of these strata, with Universal Kriging with CMAQ as only covariate yielding the best results.  In addition, all of the spatial statistical methods generate predictions with accurate uncertainty quantification: the 95\% prediction intervals constructed using spatial statistical methods have coverage at or near the nominal 95\% level.  

\begin{table}[t]\caption{{\bf Cross-validation results by the number of nearby stations:} Root mean squared error (RMSE) for $\PM$ concentration predictions in $\mu g/m^3$ stratified by the number of active monitoring sites within 50 miles of the prediction site for each day. Methods considered include: ordinary least squares (``OLS''), inverse distance weighting (``IDW''), universal Kriging (``UK''), downscaler, random forests (``RF''), support vector regression (``SVR'') and Neural networks (``NN'').  Methods include either CMAQ and/or other geographic covariates (``Covs'').}\label{t:nclose}
\centering
\begin{tabular}{l|cccc}
Active nearby stations & $<$5  & 5-9  & 10-19 & $>$=20  \\ 
  \hline
  OLS (CMAQ)        & 5.08 & 4.49 & 4.26 & 3.70 \\ 
  OLS (Covs)        & 4.88 & 4.22 & 4.34 & 3.84 \\ 
  OLS (CMAQ+Covs) & 4.53 & 3.73 & 3.77 & 3.08 \\ 
  IDW 				& 3.63 & 2.53 & 2.43 & 2.53 \\ 
  UK (CMAQ)         & 3.46 & 2.46 & 2.36 & 2.24 \\ 
  UK (Covs)         & 3.66 & 2.56 & 2.53 & 2.38 \\ 
  UK (CMAQ+Covs)  & 3.54 & 2.51 & 2.43 & 2.28 \\ 
  Downscaler(CMAQ) & 3.51 & 2.42 & 2.31 & 2.22 \\ 
  RF (CMAQ + Covs)  & 4.54 & 3.86 & 3.64 & 3.20 \\ 
  SVM (CMAQ + Covs) & 4.36 & 3.58 & 3.62 & 3.56 \\ 
  NN (CMAQ + covs)  & 4.24 & 3.45 & 3.20 & 3.11 \\ 
\end{tabular}
\end{table}

\begin{table}[ht]\caption{{\bf Cross-validation results by urbanization:} Root mean squared error (RMSE) for $\PM$ concentration predictions in $\mu g/m^3$ stratified by urbanization. Methods considered include: ordinary least squares (``OLS''), inverse distance weighting (``IDW''), universal Kriging (``UK''), downscaler, random forests (``RF''), support vector regression (``SVR'') and Neural networks (``NN'').  Methods include either CMAQ and/or other geographic covariates (``Covs''). The last line reports the mean and standard deviation, in parenthesis, for the observed PM$_{2.5}$ concentration at urban versus non-urban monitoring sites.}\label{tab:by_urbanization}
\centering
\begin{tabular}{l|cc}
  \hline
Method & Urban & Non-Urban \\ 
  \hline
  OLS (CMAQ)  & 4.74 & 5.06 \\ 
  OLS (Covs)  & 4.62 & 4.69 \\ 
  OLS (CMAQ+Covs) & 4.16 & 4.48 \\ 
  IDW & 3.08 & 3.84 \\ 
  UK (CMAQ) & 2.96 & 3.60 \\ 
  UK (Covs) & 3.16 & 3.69 \\ 
  UK (CMAQ+Covs) & 3.04 & 3.65 \\ 
  Downscaler (CMAQ) & 2.99 & 3.62 \\ 
  RF (CMAQ + Covs) & 4.18 & 4.47 \\ 
  SVM (CMAQ + Covs) & 3.78 & 4.06 \\ 
  NN (CMAQ + Covs) & 3.84 & 4.17 \\ 
\hline
  PM$_{2.5}$ & 10.16 (6.21) & 9.03 (6.01) \\
   \hline
\end{tabular}
\end{table}

\begin{table}[ht]\caption{{\bf Cross-validation results by level of observed $\PM$ concentration:} Root mean squared error (RMSE) for $\PM$ concentration predictions in $\mu g/m^3$ stratified by $\PM$ concentration level. Groupings are based on: whether $\PM$ is less than 6 $\mu g/m^3$, 6-12 $\mu g/m^3$, and greater than or equal to 12 $\mu g/m^3$ (12$\mu g/m^3$ is the EPA standard). Methods considered include: ordinary least squares (``OLS''), inverse distance weighting (``IDW''), universal Kriging (``UK''), downscaler, random forests (``RF''), support vector regression (``SVR'') and Neural networks (``NN'').  Methods include either CMAQ and/or other geographic covariates (``Covs''). The last line reports the mean and standard deviation, in parenthesis, for the observed PM$_{2.5}$ concentration within each category (low, medium or high).}\label{tab:by_magnitude}
\centering
\begin{tabular}{l|ccc}
  \hline
Method & Low & Med & High \\ 
  \hline
  OLS (CMAQ) & 2.99 & 2.63 & 7.75 \\ 
  OLS (Covs) & 2.88 & 2.76 & 7.36 \\ 
  OLS (CMAQ+Covs) & 2.46 & 2.57 & 6.73 \\ 
  IDW & 2.18 & 2.23 & 4.87 \\ 
  UK (CMAQ) & 1.72 & 1.90 & 4.93 \\ 
  UK (Covs) & 1.92 & 2.24 & 5.03 \\ 
  UK (CMAQ+Covs) & 1.79 & 1.98 & 5.02 \\ 
  Downscaler (CMAQ) & 1.77 & 1.89 & 4.97 \\ 
  RF (CMAQ + Covs) & 3.05 & 2.61 & 6.51 \\ 
  SVM (CMAQ + Covs) & 2.04 & 2.00 & 6.36 \\ 
  NN (CMAQ + Covs) & 2.62 & 2.90 & 5.75 \\ 
\hline
  PM$_{2.5}$ & 4.14 (1.21) & 8.66 (1.71) & 17.48 (5.81) \\
   \hline
\end{tabular}
\end{table}

\begin{figure}\centering
\caption{{\bf Predicted log PM$_{2.5}$ concentration for January 1, 2013} using inverse distance weighting (``IDW''), Universal Kriging with CMAQ as only predictor, the downscaler model and Random Forest. Points displayed in the map represent active monitors.}\label{f:logPMday1}
\includegraphics[page=1,width=0.48\textwidth,trim={0 3cm 0 3cm},clip]{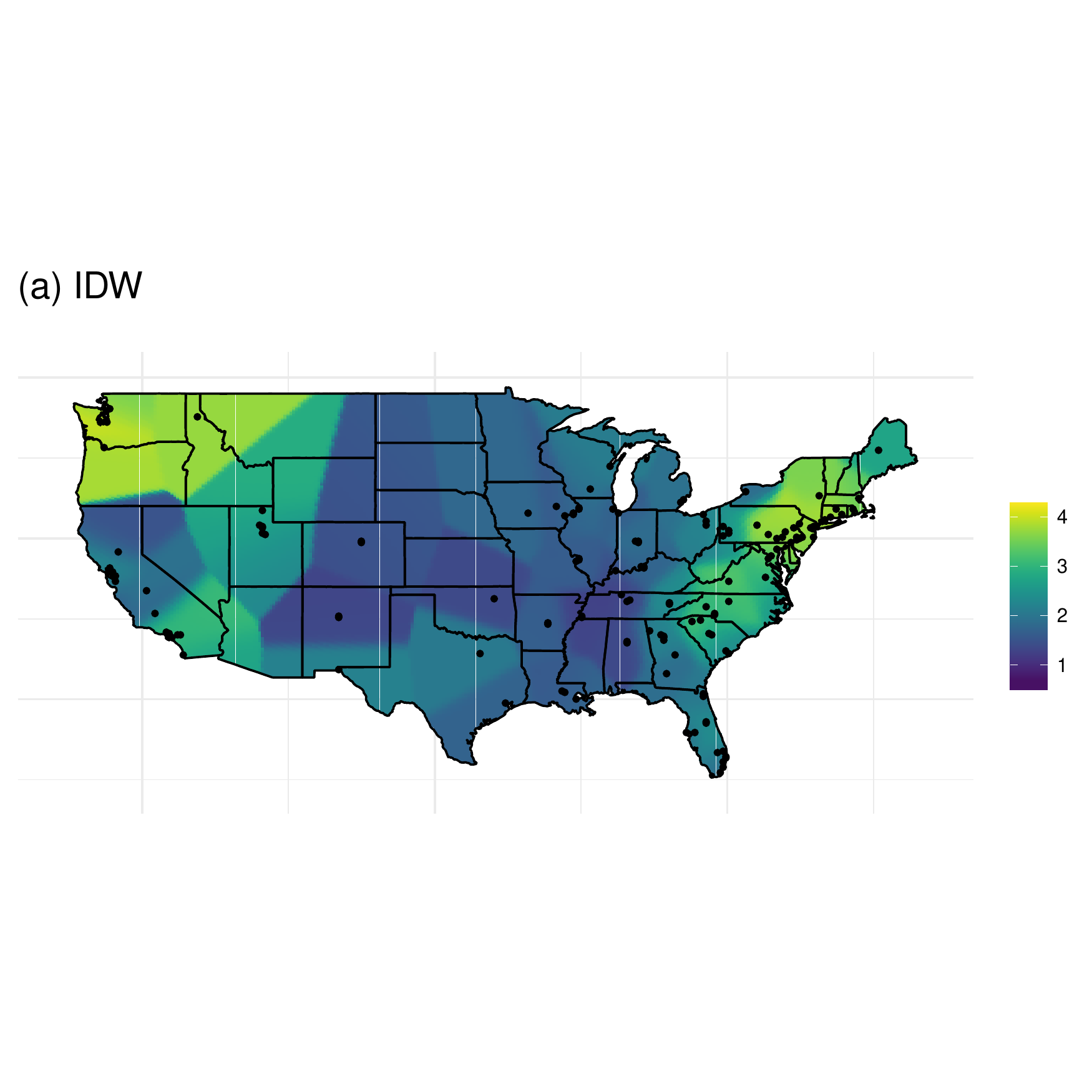}
\includegraphics[page=2,width=0.48\textwidth,trim={0 3cm 0 3cm},clip]{figs/pred_log_TRUE_day_1.pdf}\\
\includegraphics[page=3,width=0.48\textwidth,trim={0 3cm 0 3cm},clip]{figs/pred_log_TRUE_day_1.pdf}
\includegraphics[page=4,width=0.48\textwidth,trim={0 3cm 0 3cm},clip]{figs/pred_log_TRUE_day_1.pdf}
\end{figure}

\begin{figure}\centering
\caption{{\bf Predicted log PM$_{2.5}$ concentration for August 1, 2013} using inverse distance weighting (``IDW''), Universal Kriging with CMAQ as a predictor, the downscaler model and Random Forest.  Points displayed in the map represent active monitors.}\label{f:logPMday213}
\includegraphics[page=1,width=0.48\textwidth,trim={0 3cm 0 3cm},clip]{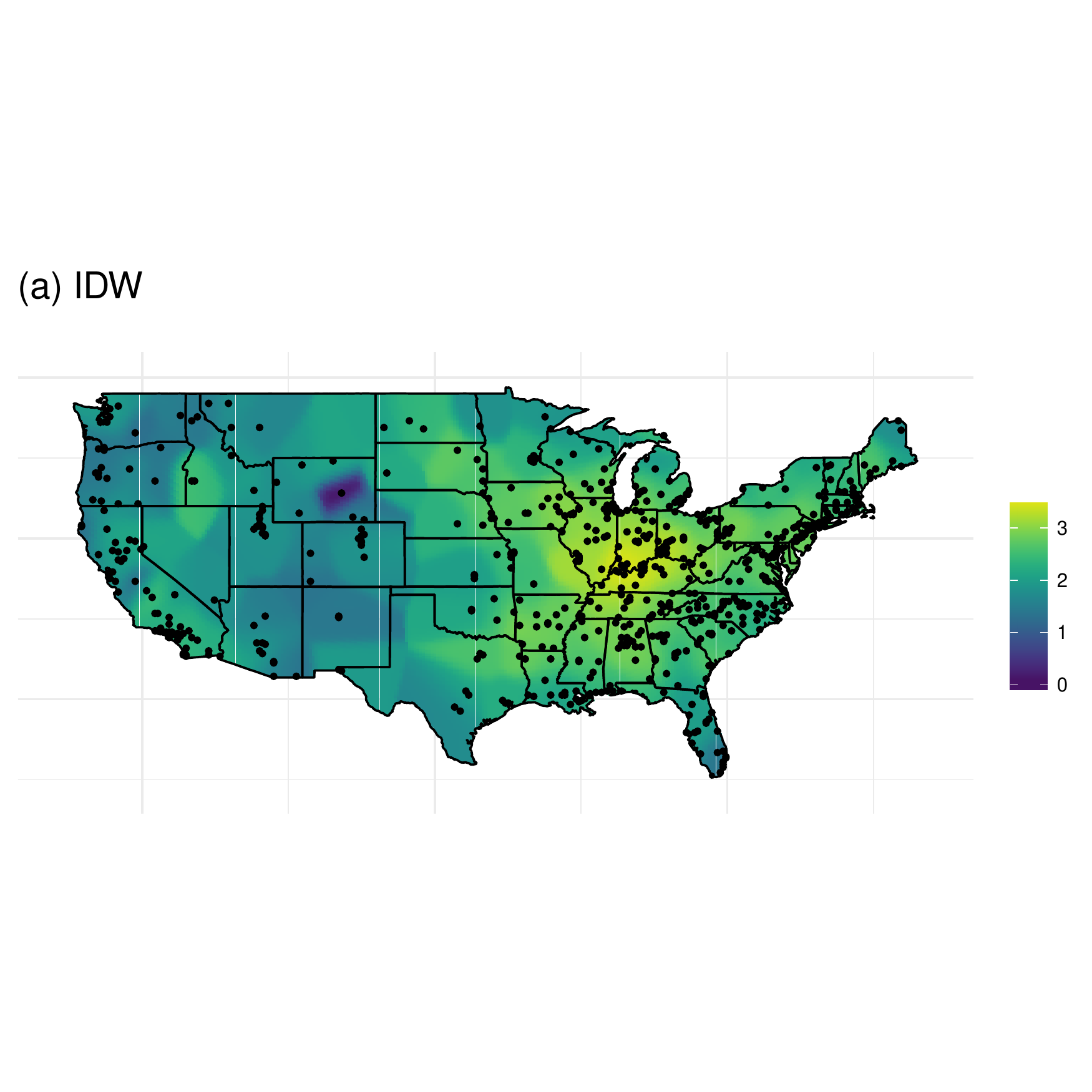}
\includegraphics[page=2,width=0.48\textwidth,trim={0 3cm 0 3cm},clip]{figs/pred_log_TRUE_day_213.pdf}\\
\includegraphics[page=3,width=0.48\textwidth,trim={0 3cm 0 3cm},clip]{figs/pred_log_TRUE_day_213.pdf}
\includegraphics[page=4,width=0.48\textwidth,trim={0 3cm 0 3cm},clip]{figs/pred_log_TRUE_day_213.pdf}
\end{figure}

Spatial predictions of log PM$_{2.5}$ over the entire contiguous United States can be seen in Figures \ref{f:logPMday1} and \ref{f:logPMday213} for the dates of January 1, 2011 and August 1, 2011, respectively. As the figures indicate, PM$_{2.5}$ concentration tends to be higher in the winter (January 1) than in the summer (August 1), with areas in the North East and the North West, especially along the coasts, having higher PM$_{2.5}$ levels. In contrast, on August 1, it is the interior region in the Eastern United States that experienced higher levels of PM$_{2.5}$ compared to the rest of the country.

The predictive surfaces of $\PM$ concentration yielded by Universal Kriging and the downscaler model present similar spatial gradients on both days. In both cases, the surfaces are generally smooth, but presents fine spatial variation particularly in areas where more monitoring data is available. Although geographical space is utilized as another independent variable in Random Forest (through the use of latitude and longitude as covariates), the spatial maps of $\PM$ concentration yielded by Random Forest are typically smooth and tend to display similar patterns to those observed in the predictive surfaces generated by the spatial statistical methods (UK and the downscaler). However, Random Forest's predictive surfaces generally have less variability than the predictive surfaces obtained using spatial statistics methods, and tend to generate predictive distributions that do not have long tails as the predicted PM$_{2.5}$ values. As a result, predictions tend to shrink closer to the mean. Finally, while the IDW method seems to yield a spatial surface that is closer to that obtained using the other methods on August 1, the predictive surface is overly smooth and does not show fine-scale spatial variability. In particular, since IDW generates predictions using only the information available from nearby monitors on each day, on days where few observations are available the predictions tend to be the same across large subregions. This leads to maps with a blocky appearance, and sudden, unrealistic, sharp changes in regions where more monitoring data is available. As such, predictive $\PM$ maps generated via IDW are not recommended as maps capturing the spatial distribution of $\PM$ concentration in the contiguous United States.

\section{Discussion}\label{s:disc}
In this paper, we present a review of commonly used statistical and machine learning methods for air pollution exposure assessment with the goal of improving our understanding of the fundamental differences among these methods. The contribution of this paper is to provide a rigorous comparison of these exposure modeling methods and make analytic code available to the broader scientific community to implement these techniques through a GitHub repository \url{https://github.com/yawenguan/DataFusion}. In our case study on $\PM$ concentration in the contiguous United States, we have found that the downscaler model and Universal Kriging yield better predictive performance than machine learning algorithms, potentially due to the fact that these methods explicitly account for the spatial dependence in $\PM$ concentration. This may hold true for other environmental exposure (e.g. ozone, or other pollutants) where spatial dependence is not negligible. 

Machine learning algorithms are attractive methods for analyzing large data sets due to their computational speed and easy implementation for massive data, partly driven by the recent availability of highly optimized computing software. In this review paper, we have chosen Random Forest, Support Vector Regression and Neural Network for comparison, because these methods have already been used for exposure modeling \citep{hu2017estimating,liu2017urban,reid2015spatiotemporal} and software within \texttt{R} is readily available. 

In our case study, we have found that the machine learning methods yield worse predictive performance than the statistical methods that account for spatial dependence. An explanation for this finding can be due to the number of variables used in our case study, which is smaller than the number of variables typically used in applications of these methods. Another reason for the underperformance of machine learning techniques might be due to the fact that these methods do not account explicitly for spatial dependence, which we have shown to be important for exposure modeling. For instance, Table \ref{tab:bydistance} compares RMSE cross-validation results of daily $\PM$ predictions by distance to the  closest active monitoring sites. When there are monitoring stations nearby the prediction site ($<$50 miles), the RMSE from the best machine learning method is worse than the best spatial statistical model: 3.61$\mu g/m^3$ for the best machine learning method versus 2.84$\mu g/m^3$ for the best spatial statistical model (ratio=1.27); on the other hand, when the monitoring stations are far ($\ge$50 miles) from the prediction site, the RMSE from the best machine learning method is 4.72$\mu g/m^3$, which is comparable to the RMSE of 4.25$\mu g/m^3$ for the best geostatistical model (ratio=1.08). The better performance by spatial statistical models in the first case is likely due to the borrowing of strength from neighboring observations. This also indicates that to improve the predictive performance of machine learning methods, algorithms that leverage values from neighboring sites for prediction, such as it is done in convolution neural net, might be useful for this application. 

\section*{Acknowledgments}
The authors thank Drs. Xuefie Hu and Yang Liu for providing the land use and meteorology data, and James Mulholland and Niru Senthilkumar for providing the CMAQ and air quality data.  This material was based upon work partially supported by the National Science Foundation under Grant DMS-1638521 to the Statistical and Applied Mathematical Sciences Institute. BJR is also supported by NIH R01ES027892, DOI-JFSP 14-1-04-9 and KAUST 3800.2; VJB is supported by NIH P30ES017885. Any opinions, findings, and conclusions or recommendations expressed in this material are those of the authors and do not necessarily reflect the views of the National Intitutes of Health and the National Science Foundation, U.S.

\section*{Appendix}
This Appendix presents maps with spatial predictions of $\PM$ concentrations in $\mu g/m^3$ over the entire contiguous United States on January 1, 2011 and August 1, 2011, respectively. In addition, it presents a comparison of the various methods for $\PM$ concentration estimation, similar to those discussed in Section 4. 

Specifically, Table 6 provides a simplified view on characteristics of the various methods compared in this case study with respect to computational speed, prediction accuracy, implementation difficulty and whether prediction uncertainty can be obtained for the various statistical and machine learning methods. We note that the computational speed varies across machines and it depends heavily on the software and code optimization. Hence, the computational speed here only serves as a rough guideline of the computational complexity and the comparisons are made based on the authors' experience. 

Additional tables compare the quality of the predictions by the various methods with stratification by distance to nearest monitoring location (Table 7), longitude (Table 8), and season (Table 9). 

\begin{table}[hb]\caption{{\bf Evaluation of downscaler methods:} The table reports whether for out data analysis the methods were executed in less than a few hours on a standard PC (``Fast''), whether they ranked near the best cross-validation error (``Accurate''), whether they were fit with standard software without many tuning parameters (``Easy'') and whether predictions were paired with measures of uncertainty (``Uncertainty''). 
}\label{t:compare}	
\begin{center}\begin{tabular}{l|cccc}
Method & Fast & Accurate & Easy & Uncertainty\\\hline 
Least squares                      & $\checkmark$ & X & $\checkmark$ & $\checkmark$ \\
Geographically-weighted regression & $\checkmark$ & $\checkmark$ & $\checkmark$ & X \\
Universal Kriging                  & $\checkmark$ & $\checkmark$ & $\checkmark$ & $\checkmark$ \\
Downscaler                         & X  & $\checkmark$ & X & $\checkmark$  \\
Random forests                     & $\checkmark$ & $\checkmark$ & $\checkmark$ & $\checkmark$ \\
Support vector regression          & $\checkmark$ & X & $\checkmark$ & X \\
Neural networks                    & $\checkmark$  & X & X & X
\end{tabular}\end{center}\end{table}

\begin{table}[ht]\caption{\textbf{RMSE Cross-validation results by distance to closest other ACTIVE monitoring station on each day.} Methods considered include: ordinary least squares (``OLS''), inverse distance weighting (``IDW''), universal Kriging (``UK''), downscaler, random forests (``RF''), support vector regression (``SVR'') and Neural networks (``NN'').  Methods include either CMAQ and/or other geographic covariates (``Covs''). The last line in the table reports the mean and standard deviation, in parenthesis, of PM$_{2.5}$ concentration observed at monitoring sites within each substratum.}\label{tab:bydistance}
\centering
\begin{tabular}{l|cc}
  \hline
Methods & Closest Station $<$ 50 Miles & Closest Station $>=$ 50 Miles \\ 
  \hline
OLS (CMAQ) & 4.67 & 5.50 \\ 
  OLS (CMAQ+Covs) & 4.06 & 5.05 \\ 
  OLS (Covs)  & 4.48 & 5.46 \\ 
  IDW & 2.95 & 4.55 \\ 
  UK (CMAQ) & 2.84 & 4.25 \\ 
  UK (CMAQ+Covs) & 2.92 & 4.33 \\ 
  UK (Covs) & 2.95 & 4.72 \\ 
  Downscaler & 2.84 & 4.36 \\ 
  RF (CMAQ + Covs) & 4.09 & 5.03 \\ 
  SVM (CMAQ + Covs) & 3.61 & 4.99 \\ 
  NN (CMAQ + Covs) & 3.74 & 4.72 \\ 
\hline
  PM$_{2.5}$ & 10.07 (6.12) & 9.36 (6.56)\\
   \hline
\end{tabular}
\end{table}

\begin{table}[ht]\caption{\textbf{RMSE Cross-validation results by East/West location in the US identified by the vertical line  longitude=-100.} Methods considered include: ordinary least squares (``OLS''), inverse distance weighting (``IDW''), universal Kriging (``UK''), downscaler, random forests (``RF''), support vector regression (``SVR'') and Neural networks (``NN'').  Methods include either CMAQ and/or other geographic covariates (``Covs'').  The last line in the table provides the mean and standard deviation (in parenthesis) for PM$_{2.5}$ concentration observed at monitoring sites within each category.}
\centering
\begin{tabular}{l|cc}
  \hline
Methods & West & East \\ 
  \hline
  OLS (CMAQ) & 7.11 & 4.80 \\ 
  OLS (Covs) & 6.47 & 4.63 \\ 
  OLS (CMAQ+Covs) & 6.20 & 4.22 \\ 
  IDW & 4.70 & 3.22 \\ 
  UK (CMAQ) & 4.74 & 3.08 \\ 
  UK (Covs) & 4.87 & 3.25 \\ 
  UK (CMAQ+Covs) & 4.84 & 3.15 \\ 
  Downscaler (CMAQ) & 4.75 & 3.10 \\ 
  RF (CMAQ + Covs) & 6.07 & 4.23 \\ 
  SVM (CMAQ + Covs)  & 5.68 & 3.83 \\ 
  NN (CMAQ + Covs) & 5.53 & 3.90 \\ 
\hline
  PM$_{2.5}$ & 8.87 (7.69) & 9.97 (6.19) \\
   \hline
\end{tabular}
\end{table}

\begin{table}[ht]\caption{\textbf{RMSE Cross-validation results by season.} Methods considered include: ordinary least squares (``OLS''), inverse distance weighting (``IDW''), universal Kriging (``UK''), downscaler, random forests (``RF''), support vector regression (``SVR'') and Neural networks (``NN'').  Methods include either CMAQ and/or other geographic covariates (``Covs'').  The last line in the table provides the mean and standard deviation, in parenthesis, for monitored PM$_{2.5}$ concentration during each season.}
\centering
\begin{tabular}{l|cccc}
  \hline
Methods & Winter & Spring & Summer & Fall \\ 
  \hline
OLS (CMAQ) & 5.93 & 4.16 & 4.37 & 4.60 \\ 
OLS (Covs) & 5.58 & 3.89 & 4.44 & 4.50 \\ 
  OLS (CMAQ+Covs) & 5.15 & 3.49 & 3.95 & 4.15 \\ 
  IDW & 3.95 & 2.69 & 2.75 & 3.38 \\ 
  UK (CMAQ) & 3.81 & 2.52 & 2.64 & 3.23 \\ 
  UK (Covs) & 3.96 & 2.60 & 2.74 & 3.57 \\ 
  UK (CMAQ+Covs) & 3.88 & 2.58 & 2.72 & 3.31 \\ 
  Downscaler (CMAQ) & 3.82 & 2.55 & 2.64 & 3.27 \\ 
  RF (CMAQ + Covs)& 5.10 & 3.52 & 4.07 & 4.15 \\ 
  SVM (CMAQ + Covs)  & 4.75 & 3.28 & 3.49 & 3.71 \\ 
  NN (CMAQ + Covs) & 4.67 & 3.21 & 3.47 & 4.13 \\ 
\hline
  PM$_{2.5}$ & 10.59 (7.05) & 9.47 (5.81) & 10.82 (5.93) & 9.01 (5.74)\\
   \hline
\end{tabular}
\end{table}

\begin{figure}\centering
\caption{{\bf Predicted PM$_{2.5}$ concentration for January 1, 2013} using inverse distance weighting (``IDW''), Universal Kriging with CMAQ as a predictor, the downscaler model and Random Forest.  Points displayed in the map represent active monitors.}\label{f:PMday1}
\includegraphics[page=1,width=0.48\textwidth,trim={0 3cm 0 3cm},clip]{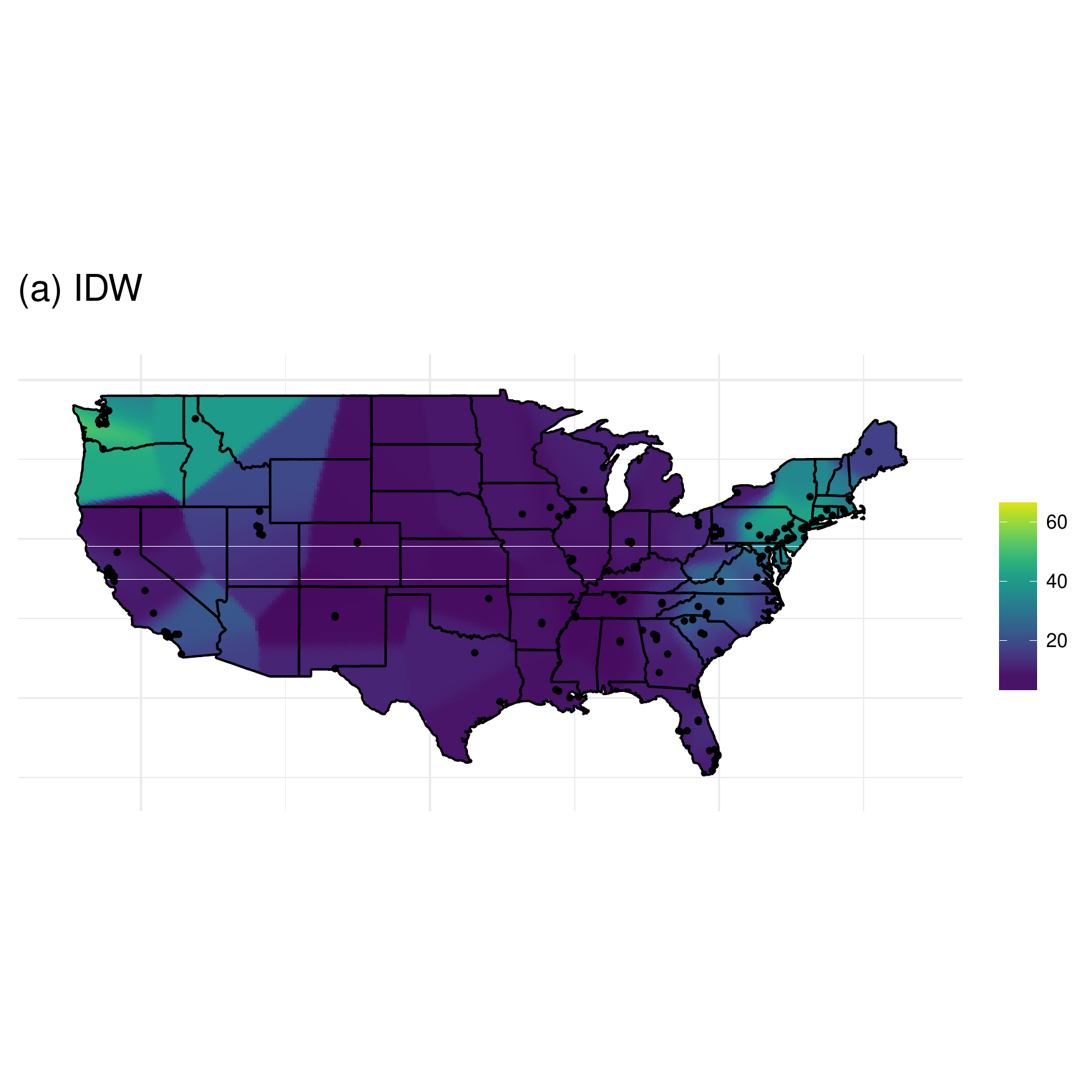}
\includegraphics[page=2,width=0.48\textwidth,trim={0 3cm 0 3cm},clip]{figs/pred_log_FALSE_day_1.pdf}\\
\includegraphics[page=3,width=0.48\textwidth,trim={0 3cm 0 3cm},clip]{figs/pred_log_FALSE_day_1.pdf}
\includegraphics[page=4,width=0.48\textwidth,trim={0 3cm 0 3cm},clip]{figs/pred_log_FALSE_day_1.pdf}
\end{figure}

\begin{figure}\centering
\caption{{\bf Predicted PM$_{2.5}$ concentration for August 1, 2013} using inverse distance weighting (``IDW''), Universal Kriging with CMAQ as a predictor, the downscaler model and Random Forest.  Points displayed in the map represent active monitors.}\label{f:PMday213}
\includegraphics[page=1,width=0.48\textwidth,trim={0 3cm 0 3cm},clip]{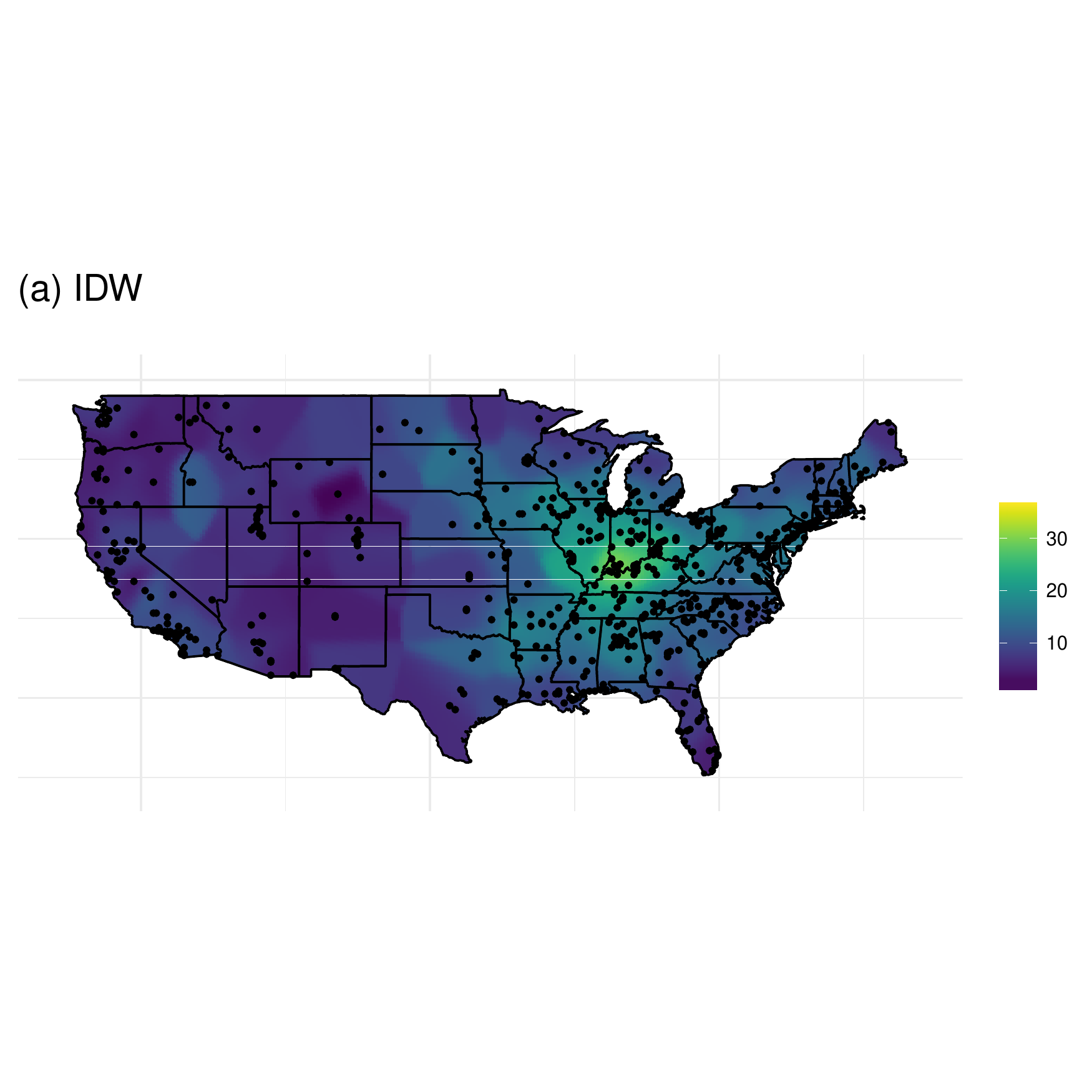}
\includegraphics[page=2,width=0.48\textwidth,trim={0 3cm 0 3cm},clip]{figs/pred_log_FALSE_day_213.pdf}\\
\includegraphics[page=3,width=0.48\textwidth,trim={0 3cm 0 3cm},clip]{figs/pred_log_FALSE_day_213.pdf}
\includegraphics[page=4,width=0.48\textwidth,trim={0 3cm 0 3cm},clip]{figs/pred_log_FALSE_day_213.pdf}
\end{figure}

\begin{singlespace}
	\bibliographystyle{rss}
	\bibliography{samsi_data_fusion} 
\end{singlespace}

\end{document}